\documentclass[runningheads,a4paper]{llncs}
\usepackage{graphicx}
\usepackage{float}
\usepackage{algorithmic}  
\usepackage[ruled,vlined]{algorithm2e}

\newcommand{\keywords}[1]{\par\addvspace\baselineskip
\noindent\keywordname\enspace\ignorespaces#1}

\newcommand{\TITLE}[1]{\item[#1]}
\newbox\fixbox
\newcommand{\algcostA}[1]{\strut\hfill\makebox[8.0cm][l]{#1}}

\newcommand{\algcostB}[1]{\strut\hfill\makebox[7.0cm][l]{#1}}
\newcommand{\algcostC}[1]{\strut\hfill\makebox[6.0cm][l]{#1}}
\newcommand{\algcostD}[1]{\strut\hfill\makebox[7.0cm][l]{#1}}
\newcommand{\algcostE}[1]{\strut\hfill\makebox[6.0cm][l]{#1}}
\newcommand{\algcostF}[1]{\strut\hfill\makebox[5.5cm][l]{#1}}
\newcommand{\algcostG}[1]{\strut\hfill\makebox[4.7cm][l]{#1}}

\begin{document}


\title{Posting with credibility in Micro-blogging systems using Digital Signatures and Watermarks: A case study on \emph{Twitter}.}

\author{Georgios K. Pitsilis\inst{1} \and Mohamed El-Hadedy\inst{2}}
\institute{Computer Science Research, Athens, Greece\\
\email{georgios.pitsilis@gmail.com}
\and Coordinated Science Lab, Urbana, Illinois, USA\\
\email{hadedy@illinois.edu}
}

\titlerunning{ }

\maketitle

\begin{abstract}
Micro-blogs are contemporary broadcasting services, for exchanging small elements of content, including video and images. Despite its popularity, micro-blogging is not without issues. So far, various security concerns, such as: privacy and confidentiality of micro-blogging systems have attracted the interest of the scientific community. Nevertheless, in this document we refer to a security issue that is concerned with the posting and circulation of fake messages. With the aim to make micro-blogs a credible source of information, in this work we propose a protocol, as a solution, that can be easily adapted to existing services such as: \emph{Twitter} and \emph{Facebook}. With the main keys of the solution being the employment of digital signatures and watermarks, the protocol can serve postings of various types of content. Finally, the solution serves the dual purpose of proofing the fake messages as well as the repudiation of postings.
\end{abstract}


\section{Introduction}

Social micro-blogging systems are increasingly popular. Unlike the traditional blogs, they allow users to exchange small elements such as short sentences (called micro-posts), which is the major reason why they received wide adoption by the users. 
Normally, micro-posts are forwarded in emails and other distribution services around the media and are used as a means for spreading the news.
It is quite common for micro-blogs to allow the inclusion of images and video links for making the posts more engaging. Despite the privacy features offered by micro-blog services, aiming to provide full control on the content published by them, there are still serious vulnerabilities. For instance, \emph{Twitter} and \emph{Facebook} are exposed to threats related to the malicious reproduction of news, or the unauthorized re-posting of the visual material \cite{Huzzlers,Yahoo,Slatest}, with both users` and services` reputation strongly being affected \cite{Slatest}.

On the other hard, there is no mechanism on the social media for tracking any intentionally withdrawn postings of a user. The lacking of mechanisms to apply such restriction has rendered micro-blogs unsuitable for the posting of highly important statements.
In \emph{Twitter}, any Internet user can register with the service, by simply choosing a user-name for identification, along with a valid email address. Then, she can freely post short messages (up to 150 characters long), which will be publicly exposed via  her personal account page. Meanwhile, other \emph{Twitter} users can register as followers to her, so that they can refer to the posts. Followers are also permitted to re-tweet messages of their followers, in which way, a reference to the first message would re-appear in the follower's account page.

\begin{figure}[!tbp]
  \centering
    \begin{minipage}[b]{0.49\textwidth}
    \centering\includegraphics[width=6cm]{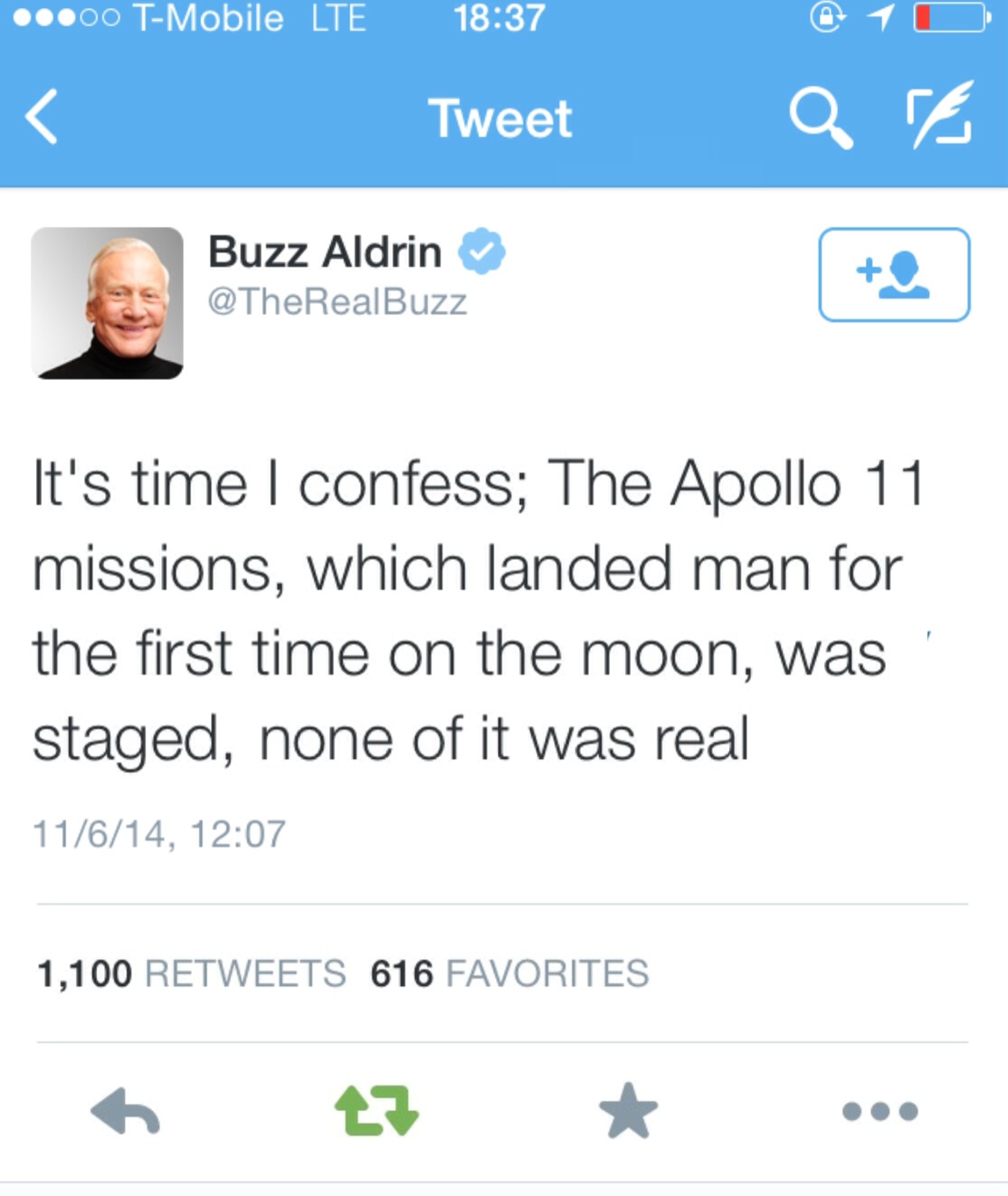}
    \caption{Example of fake tweet. (Ref. \cite{Huzzlers})}
    \label{fig:aldrin}
\end{minipage}
\hfill  
\begin{minipage}[b]{0.49\textwidth}
    \centering\includegraphics[width=6cm]{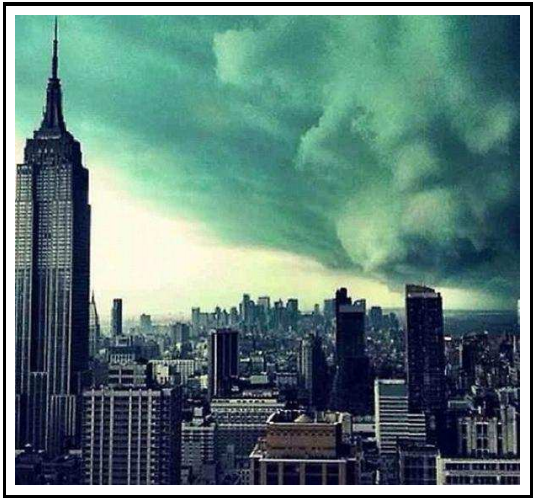}
    \caption{Example of fake picture of stormy New York skyline used in \emph{Twitter} for hurricane Sandy. (Ref. \cite{Gupta2013})}
    \label{fig:sandy}
  \end{minipage}
\end{figure}

The incentives for attacking a Micro-blogging system can be varying. The provocation of legitimate users, the acquisition of benefits from unauthorized use of the digital material, or just the spread of panic \cite{fakeTwitterBomb}, can be some of them. With respect to the technique used, attacks can be classified into: \emph{Hack tweets} and \emph{Hoax tweets}. The former regards tweets sent from a hacked legitimate account, while the latter is concerned with the posting of snapshots of fake tweets to various websites. (See example in fig.\ref{fig:aldrin}). As far as the former type, an incident known as \emph{Hack crash} \cite{HackCrash}, shows the level of damage that can be caused by such attacks. In \emph{Hack crash}, tweets were sent from a hacked \emph{Twitter} account resulted to short-time panic in the US stock market in  April 2013 \cite{panicWallStreet}. The use of automated financial algorithms in stock market systems, which retrieve information from social media to predict future trends of the market \cite{Papaioannou2013}, was mainly responsible for the panic caused. In that incident, the extensive automation in the algorithms used, as they were reacting quicker than human thought, gave no time for the involved parties to discover the nature of tweets.



As far as the \emph{Hoax tweet} type, a threat can be simply thought of as a malice case of user creating fake tweet content, claimed as being posted by the attacked account. The publishing of visuals alongside tweets, with text not tightly attached to the images, offers space for persuasive \emph{hoax-tweets}. The availability of many on-line services for creating fake tweets of any content \cite{tweetfake,simitator,lemmetweet} is a factor for encouraging attackers to target any legitimate \emph{Twitter} user. It is worth mention, the existence of various toolkits, such as \emph{Twitt-embeds} by \emph{Twitter} API \cite{embedTwitt}, which provide seamless embedding of tweets into a web page, and which could be potentially used for malicious purposes. The output of \emph{Twitt-embeds}, by just being an ordinary image file, can be accessed by any user without necessarily being a \emph{Twitter} user. That could be a potential vulnerability for the \emph{Twitter} service itself, because since it can not be resolved if an embedded tweet that looks original, is truly posted or faked.

For \emph{Facebook}, the unauthorized reposting of images already posted by legitimate users \cite{Yahoo}, is the main threat for the service itself, as there is lack of protection for media distribution. The similarities between the nature of the services and the security policies used in \emph{Facebook} and \emph{Twitter}, suggests applying some generic solution.

\section{Existing approaches}
\emph{Twitter} does not employ any security mechanism to avoid the above incidents.
For \emph{Twitter}, the only means of protection so far for the account owners, regards restrictions that can be provisionally applied to followers, allowing only those approved by the account owner to access the posted material. Nevertheless, such policy, in the way it works, has the result on tweets’ visibility seriously being affected. Other approaches include classification techniques which are known to achieve some success on detecting images that look irrelevant with the posted text on \emph{Twitter} \cite{Gupta2013}. (see fig.
\ref{fig:sandy}.)

While the application of higher security measures on the existing user authentication system could, at some point, prohibit \emph{Hack tweets}, on the other hand, the prevention of \emph{Hoax tweets} requires a different approach, which we elaborate in this document.

\section{The proposed solution}

The solution we propose focuses on the issues of \emph{Hoax-tweets} and it provides two alternative schemes that combine: Watermarking, One-way functions and Public Key encryption. The main idea is that, the posted material, prior to publishing in the micro-blogging system, would first get digitally signed and uniquely tagged with appropriate evidence to prove its origin.
 
Considering \emph{Twitter} as a pilot use case, with tweets consisting of text alone, a likely solution would be to apply hashing and pair-key cryptography. Hashing is applied onto the text content to shorten its size. This is quite necessary for messages of long size. To create a unique ID for the message, the hash value is signed with the user`s Private key. For the case of tweets comprising images, we propose the solution of embedding the signatures into the images themselves in the form of watermarks. To enhance robustness, the tweet signatures could include a digest of the attached image. 

Next, we explain in more detail the protocol that implements the proposed scheme, as well as some variations of it. 
All the schemes we propose require the existence of a Trusted Third Party (TTP), in which, users who wish to take advantage of the protection scheme, will first need to register with a Private key.
(See algorithm \ref{protocol1}.)
TTP could also serve as the entity to perform the watermarking service. 
In such scheme, the resolution, that is the verification of authenticity of some post, can be performed by any entity with the provision of the Public key of the tweet's owner.

The following symbols have been used: The entity $A$ denote as the owner of the \emph{Twitter} account and $B$ is the entity serving as TTP. $K_A$ is the Private key of $A$, while $K_{PA}$ is his Public Key. $B$ could also run on the client side ( user's hardware), without any compromise in the security.

\begin{algorithm}
\caption{Registration with TTP}
\label{protocol1}
\begin{algorithmic}[1]
\TITLE{\textsc{Communication}}
\algcostA{\textsc{Comment}}
\STATE $A \rightarrow B: K_A $ \algcostA{Compute and send Private key to trusted party $B$}
\STATE $A \rightarrow B: K_{PA}$ \algcostA{Compute and send Public key to trusted party $B$}
\end{algorithmic}  
\end{algorithm}

\subsection{For tweets containing text alone}

For tweets consisted of text alone, prior to 
publication to \emph{Twitter} service $C$, the text content $M$ would be first hashed ( $\#M$ ), before it gets signed with the private key $K_A$ of user $A$. (See algorithm \ref{protocol2}). The final tweet includes the text message (step 2), plus a unique ID, which we call the \emph{key-code} of the tweet. That code will be shown publicly, along with the posted text (step 3).

\begin{algorithm}
\caption{Tweeting text alone}  
\label{protocol2}
\begin{algorithmic}[1]
\TITLE{\textsc{Communication}}
\algcostB{\textsc{Comment}}
\STATE $A \rightarrow B: M$ \algcostB{Send text message $M$ to trusted party $B$}
\STATE $B \rightarrow C: M$ \algcostB{Forward text message $M$ to \emph{Twitter} service $C$}
\STATE $B \rightarrow C: \{\#(M)\}_{K_A}$ \algcostB{\emph{key-code} posted along the hashed tweet text $M$}
\end{algorithmic}  
\end{algorithm}

For verifying the authenticity of text tweets, we include another entity $D$ in the scheme, acting as someone wishing to verify the origin of some tweet. (See algorithm \ref{protocol3}). There is no requirement in the scheme for $D$ to be a trusted party. As such, it is reasonable to assume that $D$ should be allowed to use as input, the publicly shown text content $M$ of a tweet, along with the  \emph{key-code}  $\{\#(M)\}_{K_A}$  computed for that content. Then, by decrypting the \emph{key-code}, the signature of the hash value is received (step 2). Next, $D$ would hash the twitted plain-text $M$ to get the signature of it (step 3), which he would then compare with the output of the previous step (2), to verify the origin of the tweet.\\

\begin{algorithm}[H]
\caption{Verification of text-based tweets}  
\label{protocol3}
\begin{algorithmic}[1]
\TITLE{\textsc{Communication}}
\algcostC{\textsc{Comment}}
\STATE $B \rightarrow D: K_{PA}$ \algcostC{Public key of $A$}
\STATE $C \rightarrow D: \{\{\#(M)\}_{K_A}\}_{K_{PA}}$  \algcostC{Verifying the originality of the \emph{key-code}}
\STATE $C \rightarrow D: M$  \algcostC{by comparing the output with} 
\item[] \algcostC{the plain text $M$}
\end{algorithmic}  
\end{algorithm}

\subsection{For images posted alongside text tweets}
\label{simple}

Images posted alongside text, should be first watermarked with the signed \emph{key-code} ID of the text content, as outlined in the previous step. That is nessesary for creating a connection between the textual and the media component of a post. The process is shown in algorithm \ref{protocol4} and is detailed as follows: The expression $wm( I,P )$ in step 3 denotes the operation of hiding the string $I$ into picture $P$ by watermarking, which in our case, the picture is the image the user wishes to post alongside a text tweet. As before, the watermarking task can be carried out by the trusted entity $B$. The scheme allows the posting of more than one image alongside a text tweet, which can be achieved with the repetition of step 3 for each additional image. 

\begin{algorithm}
\caption{Attaching an image to a tweet}  
\label{protocol4}
\begin{algorithmic}[1]
\TITLE{\textsc{Communication}}
\algcostD{\textsc{Comment}}
\STATE $A \rightarrow B: \{\#(M)\}_{K_A}$ \algcostD{The \emph{key-code} is computed and submitted to}
\item[] \algcostD{the trusted party $B$}
\STATE $A \rightarrow B$: $P$ \algcostD{The picture $P$ is sent to the trusted party $B$}
\STATE $B \rightarrow C$: $wm( \{\#(M)\}_{K_A},P )$ \algcostD{Embedding the encrypted hash value of $M$}
\item[] \algcostD{into the watermarked picture}
\STATE $B \rightarrow C$ : $\{\#(M)\}_{K_A}$ \algcostD{The \emph{key-code} is posted to \emph{Twitter} service $C$}
\item[] \algcostD{alongside the watermarked picture}
\STATE $B \rightarrow C$: $M$ \algcostD{The text is finally posted to \emph{Twitter} service $C$}
\end{algorithmic}  
\end{algorithm}

In the verification shown in algorithm \ref{protocol5}, the expression $exw (P_{pub},V)$  denote as the function used for extracting a watermark with content $V$ from the watermarked picture $P_{pub}$. This function returns either $V$ or null in case the content $V$ has not been detected into $P_{pub}$.
The assumption made is that, extracting a watermark actually means detecting the existence of a particular sequence of bytes within an image. This has been found to be a good practice for watermarks used for serving a security purpose \cite{HadedyPitsilis2011}.
Therefore, the verification scheme we present can only serve the case when only requiring to answer the question, whether a picture has been posted along with a text tweet by some user $X$ or not. Such verification scheme also provides the benefit of not needing to store the original, non-watermarked picture somewhere in the system, thus enhancing the security of the scheme.

\begin{algorithm}
\caption{Verification of pictured tweets}  
\label{protocol5}
\begin{algorithmic}[1]
\TITLE{\textsc{Communication}}
\algcostE{\textsc{Comment}}
\STATE $C \rightarrow D$ : $M$ \algcostE{The message in plain text as posted}
\STATE $B \rightarrow D$ : $K_{PA}$ \algcostE{The Public key for decryption}
\STATE $C \rightarrow D$ : $\{\#(M)\}_{K_A}$  \algcostE{The \emph{key-code} of the \emph{Twitter} posting}
\STATE $C \rightarrow D$ : $\{exw(P_{pub},\#(M)_{K_A})\}_{K_{AP}}$ \algcostE{Extracting and decrypting the info}
\item[] \algcostE{to verify the \emph{key-code}}
\end{algorithmic}
\end{algorithm}

\subsection{Improved scheme for higher robustness}

For improved security we propose an alternative protocol, which provides over the simplified version in section \ref{simple}, a secure link between the tweeted text and any pictures attached to it. In addition to the simplified version, the tweet`s \emph{key-code} ID is derived from both the image content and the text tweet. Furthermore, a time-stamp has been included in the \emph{key-code}. That was for securing the protocol against likely attacks in which, text and images from various tweets posted by the same legitimate user $A$ in the past, are mixed together by an attacker to produce a hoax-tweet.\\

For example, lets assume the scenario in which a legitimate user has posted a tweet at time T1, and is composed of a text message M1 and a picture P1. 
The same user, at time T2 posts another message with text content M2 that is exactly the same with M1, but this time attaching picture P2 to it. 
If algorithm \ref{protocol4} was used, an attacker would still be able to violate the security of the system, by creating a new hoax tweet at time T3, composed of the textual content T1 attached to picture P2.

Despite both text and image have been posted by the same legitimate user, an attacker can still create a hoax one, by mixing the components across tweets.
More important, the hoax tweet yet would look original, because there is no way in the former scheme to prove that such tweet can be the product of tampering. The reason is the use of unaltered components of legitimate tweets to make the hoax tweet.

Likewise, for tweets produced by their account owners, there is no way to oppose a complainant's claim that such tweets are not original. 
We attribute the reason for the above issues, to the fact that, in the above design the components of a tweet cannot be uniquely referred.\\

The general idea of an alternative protocol to tackle the above issue can be described as follows: The \emph{key-code} ID is composed of two parts, concatenated together: The first part contains the ID provided by hashing the text message with the time-stamp, while the second part can be a unique value, associated with the attached picture. 
As opposed to the simple version of the protocol in algorithm \ref{protocol4}, in this one, the time-stamp was included for ensuring the uniqueness of the composed \emph{key-code} ID. 
In this way, likely collisions can be prevented, that may occur when attempted the use of the same  \emph{key-code} in two different tweets. Furthermore, the scheme can make provable as well, if some tweet text or some picture have ever been posted before.

The protocol is detailed in algorithm \ref{protocol6} and it works as follows: A digest of the image is produced by hashing its file content, in step 3. Then, the produced unique ID is used along with the text message $M$, which is also hashed for signing the tweet. 

In the notation used, the plus symbol ( + ) denotes concatenation. $T$ is used to denote a time-stamp. For short, we call $P_{pub}=wm ([{\#(M+T)}_{K_A}],P)$ the publicly shown watermarked picture that is attached to the tweet, while P denote as the original picture.

\begin{algorithm}
\caption{Attaching images on a tweet. The provable way}  
\label{protocol6}
\begin{algorithmic}[1]
  \TITLE{\textsc{Communication}}
  \algcostF{\textsc{Comment}}
\STATE $A \rightarrow B$ : $P,M,T$ \algcostF{The message $M$, the time-stamp $T$}
\item[] \algcostF{and the picture $P$ are send over to}
\item[] \algcostF{the trusted third party $B$.}
\STATE $B \rightarrow C$ : $wm( [ \{\#(M+T)\}_{K_A}] ,P)$ \algcostF{The watermarking of time-stamped}
\item[] \algcostF{message into the picture, is}
\item[] \algcostF{performed by the trusted party.}
\STATE $B \rightarrow C$ : $\{\#(M+T)\}_{K_A}+\{ \#(P_{pub}) \}_{K_A}$ \algcostF{The concatenation of $M+T$ with}
\item[] \algcostF{the picture, provides a unique}
\item[] \algcostF{\emph{key-code} to display, composed of}
\item[] \algcostF{two segments.}
\STATE $B \rightarrow C$ : $M+T$ \algcostF{The message $M$ and time-stamp $T$}
\item[] \algcostF{are posted along with the \emph{key-code}.}
\end{algorithmic}
\end{algorithm}

In step 2, the digest of the time-stamped message content is embedded into the watermark, as a way to uniquely referring to it. In step 3, the unique tweet ID (\emph{key-code}) is derived from both the picture and the text message content. In addition, $T$ has been included to prevent attacks, in which a malicious user could create a hoax tweet by mixing various components from past tweets.

If required to attach multiple pictures to a tweet, the 3rd step can be modified accordingly to accommodate the signatures of all pictures together. As such, step 3 should be repeated for every picture to be attached.\\

\begin{algorithm}[H]
\caption{Verification of provable pictured tweets}  
\label{protocol7}
\begin{algorithmic}[1]
  \TITLE{\textsc{Communication}}
  \algcostG{\textsc{Comment}}
\STATE $B \rightarrow D$ : $K_{PA}$ \algcostG{The public key of $A$}
\STATE $C \rightarrow D$ : $M+T$    \algcostG{The message timestamped}
\STATE $C \rightarrow D$ : $\{\#(M+T)\}_{K_A}+\{ \#(P_{pub}) \}_{K_A}$ \algcostG{The \emph{key-code} of the posted}
\item[] \algcostG{tweet is provided as public}
\item[] \algcostG{information.}
\STATE $C \rightarrow D$ : $\{exw(P_{pub},\{\#(M+T)\}_{K_A})\}_{K_{PA}}$ \algcostG{Extracting, decryptinig and}
\item[] \algcostG{verifying the text message}
\item[] \algcostG{from the picture.}
\item[]
\STATE $C \rightarrow D$ : $\{\{\#(M+T)\}_{K_A}\}_{K_{PA}}+ \{\{\#(P_{pub})\}_{K_A}\}_{K_{PA}}$
\item[] \algcostG{Verifying the originality}
\item[] \algcostG{of the \emph{key-code.}}
\end{algorithmic}
\end{algorithm}

As previously mentioned, the verification serves a dual purpose. First it is to verify that some picture has been posted along with a particular tweet (step 4), and second, to prove the originality of the composed tweet as a whole (step 5). 
In step 4, the first part alone of the \emph{key-code} $\{\#(M+T)\}_{K_A}$, suffices to confirm that the picture $P$ has been truly attached by the legitimate user in that tweet.
The verification of the originality of the \emph{key-code} is performed in step 5, were used as input the watermarked picture $P_{pub}$ along with the timestamped plain message $M+T$.
In this way can be proved the originality of a tweet, by checking whether the \emph{key-code} can be succesfully re-composed from these 3 elements.
That is done by decrypting each half of the key-code in step 5, and comparing it with its adjacent part from step 3. 


\section{Implementation for the \emph{Twitter} service}

\begin{figure}[h]
\hspace*{-1cm} 
\centering\includegraphics[scale=0.35]{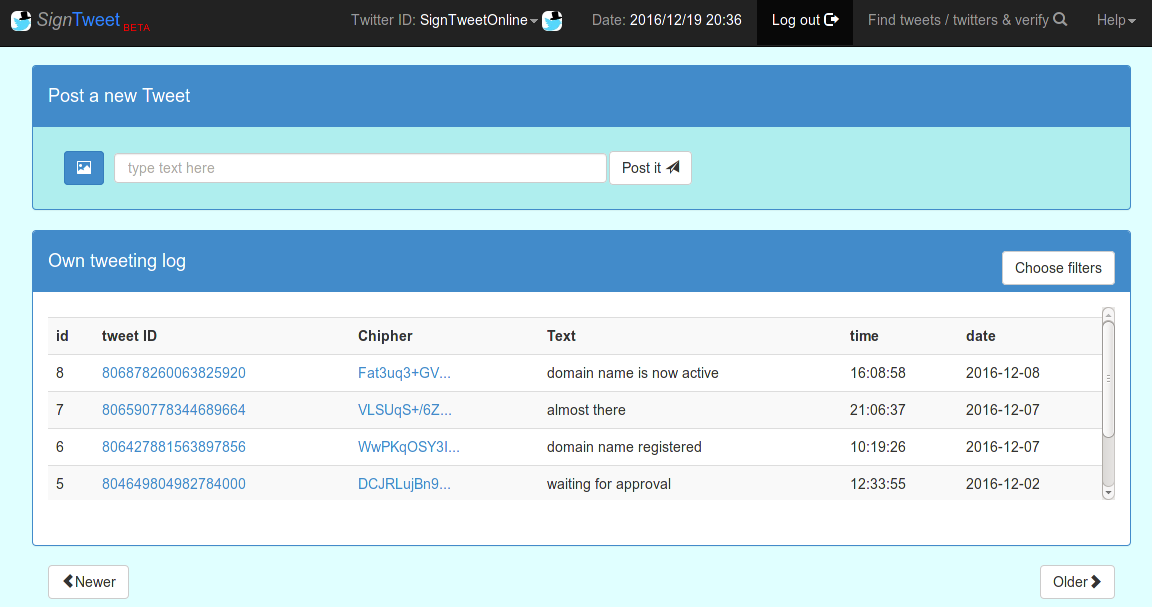}
\caption{The user interface for posting tweets}
\label{fig:signTweet1}
\end{figure}

\begin{figure}[h]
\hspace*{-0.5cm} 
\centering\includegraphics[scale=0.35]{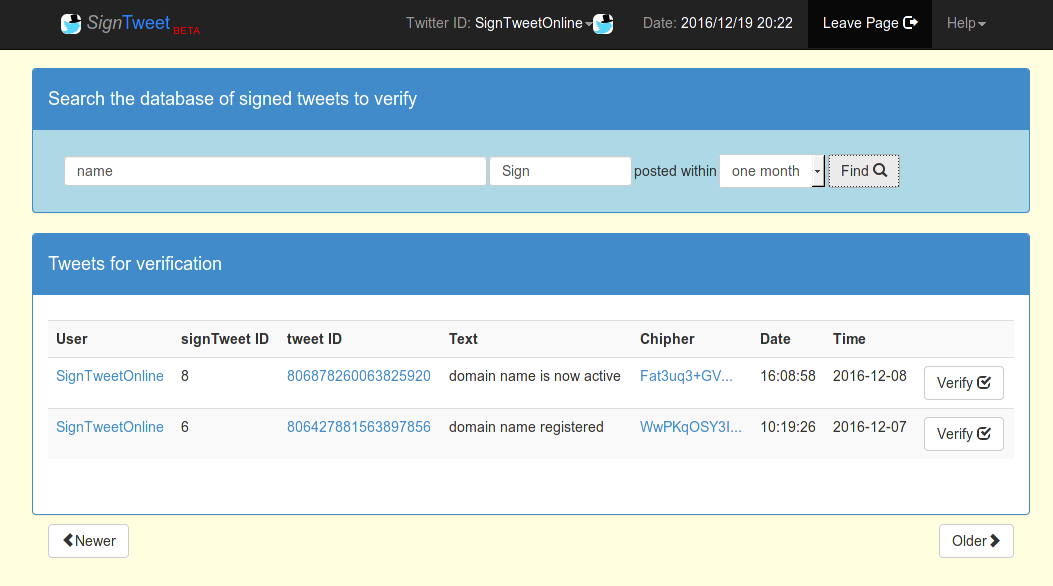}
\caption{The user interface for searching and verifying tweets.}
\label{fig:signTweet2}
\end{figure}

As a proof of concept, we implemented a service that follows the above design.
The provided UI allows to any user already having a \emph{twitter} account to use his credentials for signing up to this service and start using it. 
As such, any new tweets posted by this user via our UI, will become securely signed before they get posted on \emph{Twitter}. (See fig.\ref{fig:signTweet1})
Furthermore, all the evidence required for the verification of tweets, become publicly available in our service's web site. The service also features a publicly accessible searching and browsing tool, for finding tweets posted by any registered user. (See fig.\ref{fig:signTweet2})

The status of the current implementation is: "work in progress", therefore it currently supports only a subset of the full functionality. More particularly, only text tweets can be submitted and verified. We left the watermarking feature of any attached pictures as a future work.

We should also point out that, for this implementation, due to the short message length allowed in the \emph{Twitter} service, we found unnecessary to apply hashing onto the text messages (as per step 3 in algorithm 2). That is because hashing very short messages results to lengthy cipher-text, which is not practical for \emph{Twitter}.

The following software tools were used in the implementation :
\emph {PHP phpseclib} \cite{phpseclib}, which contains the \emph{RSA encryption library}, and the \emph{base64-encode} \& \emph{base64-decode} modules.
For accessing the users data and services offered by the \emph{Twitter} platform through program code, we used an API, especially built by Abraham \cite{abraham}. The above API was chosen as it can work with PHP server-side scripting language and functions as REST interface.

To avoid any misinterpretation of the information by the communication protocols during the transmission across the network, we used 64-base encoding for the storing of cipher-text and encryption keys in the system.

\begin{figure}[h]
\centering\includegraphics[scale=0.35]{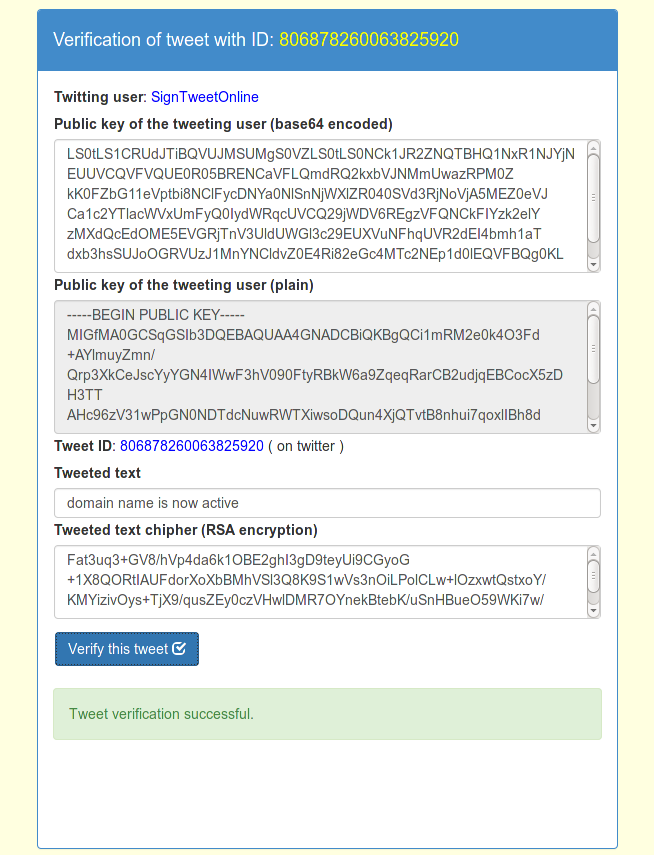}
\caption{Tweet verification result}
\label{fig:verify}
\end{figure}

To eliminate any doubts regarding the validity of the verification, as can be seen in fig. \ref{fig:verify}, the result is presented along with all the information necessary for anyone to be able to re-validate this step by other means. In that respect, the employment of external tools with same functionality should be sufficient to check the validation of the process.
For instance, by applying RSA decryption onto the \emph{Tweeted Text Cipher} using the presented \emph{Public Key} is the way to perform custom verification.
As per algorithm \ref{protocol3}, showing a decryption output that equals to the displayed \emph{Tweeted Text}, would be enough to prove the originality of the posted message.


Our service currently serves experimental purposes only and it is totally free of charge. The service can be reached at the web address: \emph{http://www.signtweet.com}.\\

Next, we present a likely scenario of use:

\begin{enumerate}
\item Registration. Users who have not been registered with our systems, can do so by using their \emph{Twitter} credentials. Upon registration, both the Private and Public keys are automatically computed by our system, and are allocated to the new user.
\item Posting a tweet. Upon submitting a text tweet using the signTweet GUI (See fig. \ref{fig:signTweet1}), a cipher-text is computed based upon the Private key, and it is presented in the user's history page, along with the plain-text. Afterwards, the textual content is automatically forwarded and displayed in the user's official page in \emph{Twitter}.
\item Verification. This step is optional, it can be run by anyone, hence not requiring registration. The purpose served is to provide the means to verify the originality of any tweets posted through our service. 
Using our UI, any tweet posted via our service can be retrieved by applying the proper criteria. (See fig. \ref{fig:signTweet2}.) Next, by clicking the \emph{Verify} button for this tweet, in a separate window is shown all the evidence required for the verification, such as: the Public key of the posting user, the cipher-text of the tweet as well as the plain text. (See fig.\ref{fig:verify}).
Finally, by clicking the button created for this purpose, and it is called "Vefiry this tweet", the system proceeds to the verification based upon the above data.
It is worth mention that, the data in the verification window are in editable form, so that the user can examine the verification output in an interactive way, by optionally modifying the evidence.
\end{enumerate}

Future work in our implementation schedule includes, the support of the protocol for other forms of posting material, such as photos, as per algorithms \ref{protocol4},\ref{protocol5},\ref{protocol6} and \ref{protocol7}.

\section{Discussion}

The proposed design can successfully serve a dual purpose as we described previously. In one use-case, it can serve the purpose of account owners wishing to prove that any claims by other users about a posted tweet is truly a hoax \cite{Huzzlers}. 
In another use-case, in which is claimed that offensive tweets have been publicly posted, but removed afterwards by the legitimate account owner, the proposed scheme can make possible the verification of such claim. 
Whoever has retained a snapshot of the tweet in question along with the cipher-text, can certainly prove his claim that such tweet has ever been posted by the legitimate account owner. In either case, the assumption is, the embedding of such service into micro-blogs, will be for the benefit of whoever is operating at good practice. (either a legitimate \emph{Twitter} account owner, or a complainant). 
In that way, potential attackers would be discouraged from producing fake tweets for any user who has registered to our service.
Similarly, a legitimate \emph{Twitter} user, by adapting such protection scheme, could easily build a reputation of behaving in good practice.

\section{Conclusion}

In this document we provided an algorithm that could help to make micro-blogging systems more credible and secure. It can support micro-posts of multiple types, such as text alone, or text with attached images. We provided algorithms for securing and verifying the authenticity of posted messages using Public Key Encryption, Digital Signatures and Watermarking. We also presented a proof-of-concept implementation of the above algorithm in the form of a publicly accessible service. We are hoping such protocol to receive great adoption by the existing micro-blogging services.

\section{Acknowledgements}

We would like to thank \emph{Prof. Kevin Skadron} from the Department of Computer Science, School of Engineering and Applied Science, University of Virginia, US, for his contribution and the useful discussions on the subject.

\bibliographystyle{abbrv}
\bibliography{st.bib}

\end{document}